\documentclass[11pt]{article}
\usepackage{blois,epsfig}

\usepackage{mathrsfs}

\def\be{\begin{equation}}
\def\ee{\end{equation}}
\def\bea{\begin{eqnarray}}
\def\eea{\end{eqnarray}}

\begin{document}
\vspace*{4cm}
\title{Directional detection of non-baryonic dark matter with MIMAC}

\author{C.~Grignon, J.~Billard,  G.~Bosson, O.~Bourrion, O.~Guillaudin, F.~Mayet, J.P.~Richer, D.~Santos}

\address{LPSC, Universite Joseph Fourier Grenoble 1, CNRS/IN2P3, Institut Polytechnique de Grenoble}

\author{E.~Ferrer, I.~Giomataris, J.P.~Mols}

\address{CEA Saclay, 91191 Gif-sur-Yvette cedex}

\maketitle\abstracts{
Directional detection of non-baryonic Dark Matter is a promising search strategy for discriminating genuine WIMP events from background ones. However, carrying out such a strategy requires both a precise measurement of the energy down to a few keV and 3D reconstruction of tracks down to a few mm. To achieve this goal, the MIMAC project has been developed: it is based on a gaseous micro-TPC matrix, filled with $\rm ^3He$, $\rm CF_4$ and/or $\rm C_4H_{10}$. Firsts results of low energy nuclei recoils obtained with a low energy neutron field are presented.
}

\section{Directional detection of Dark Matter and its discovery potential}\label{sec:principle}

In order to offer a complementarity 
to massive detectors of non-baryonic dark matter  \cite{xenon,edelweiss-armengaud}, a solid and unambiguous signature of WIMP signal is needed. 
This can be achieved by searching for a correlation of the WIMP signal with  the solar motion around the galactic center, observed as a direction dependence of the WIMP flux  coming roughly from  the direction of the Cygnus constellation \cite{spergel}. 
This is generally referred to as directional  detection of Dark Matter and several projects of detector are being developed for 
this goal \cite{MIMAC,Drift,mit,newage,white}.\\

\begin{figure}[h!]
\begin{center}
\includegraphics[scale=0.1864]{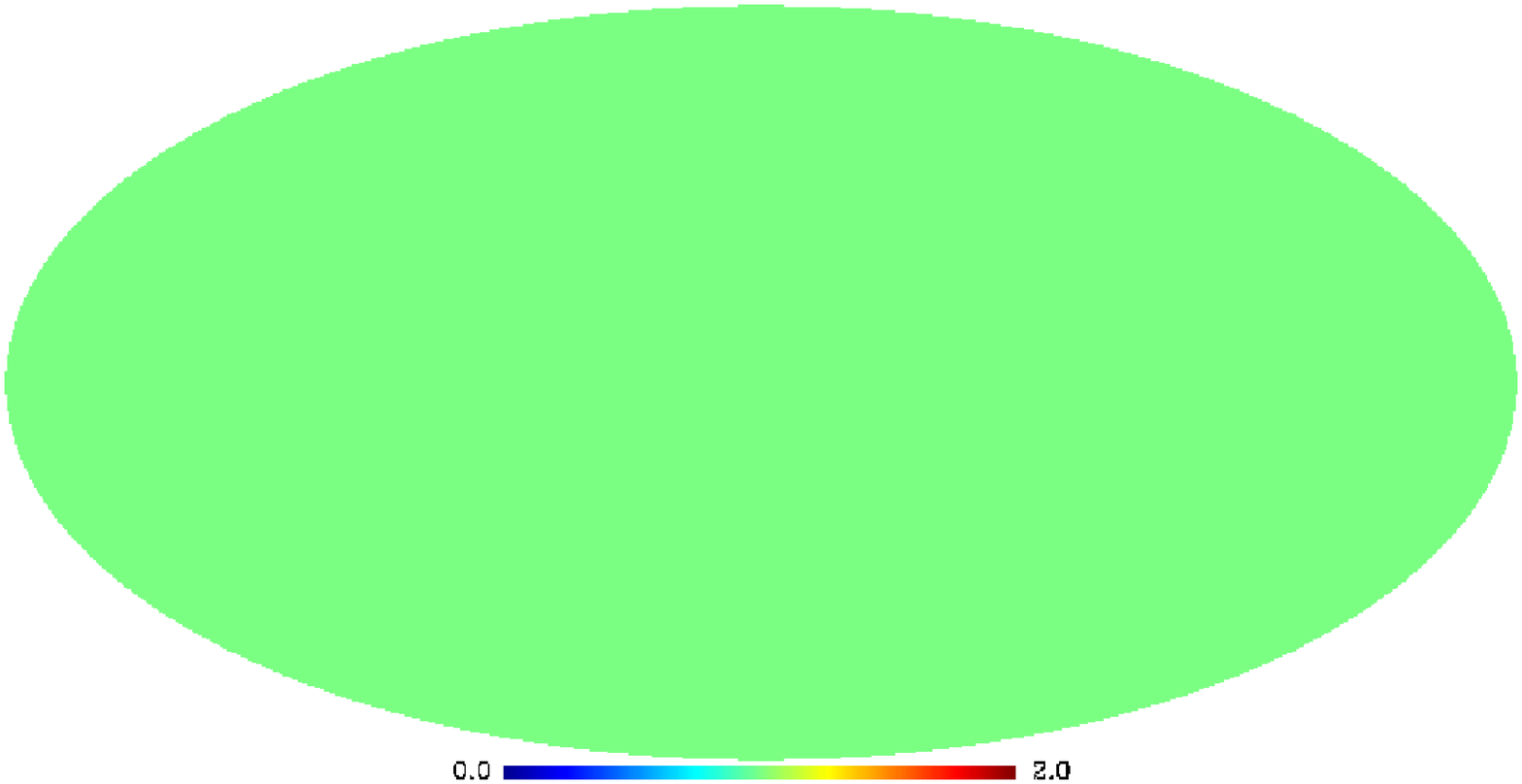}
\includegraphics[scale=0.2,angle=90]{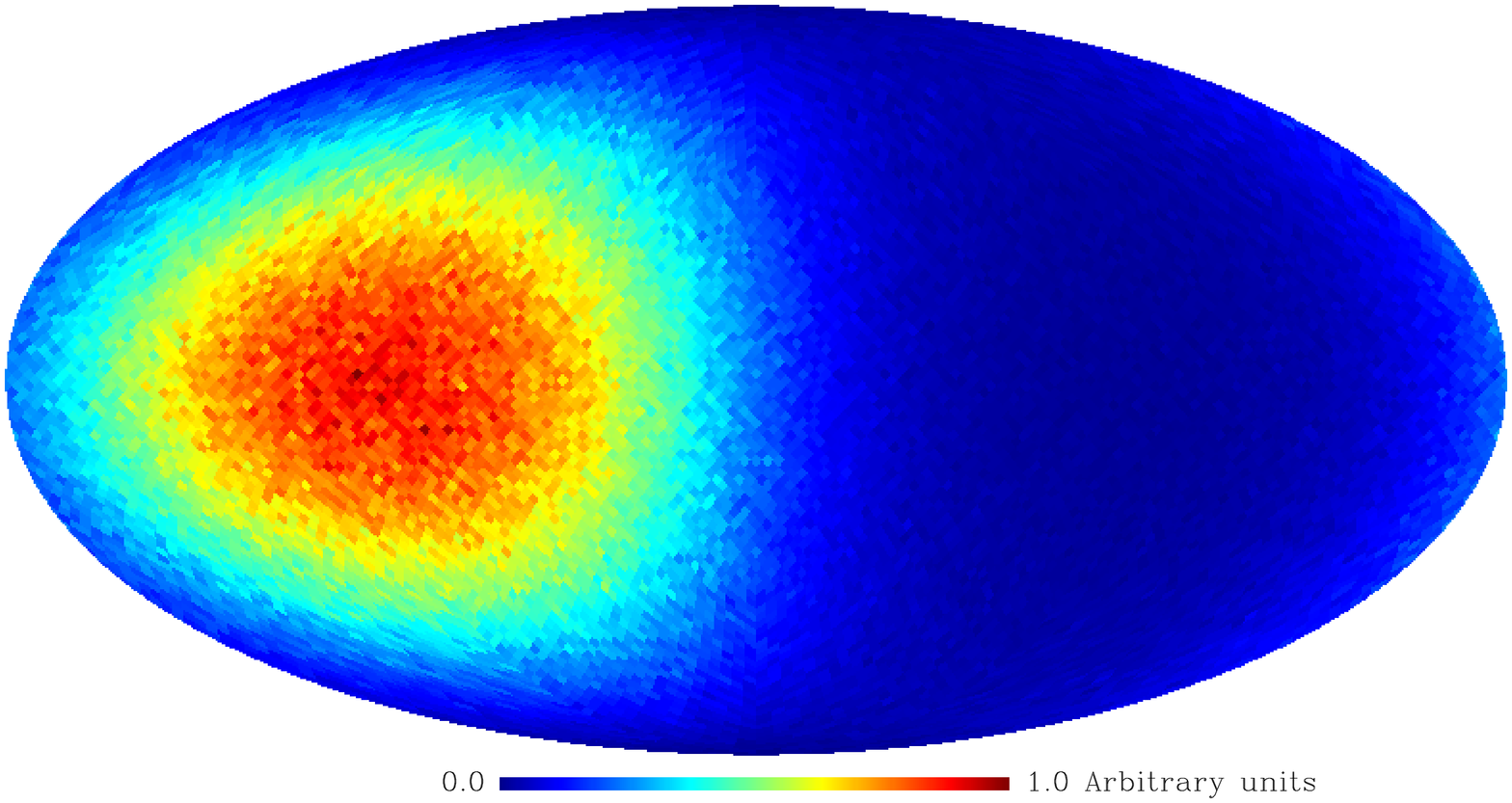}
\includegraphics[scale=0.2,angle=90]{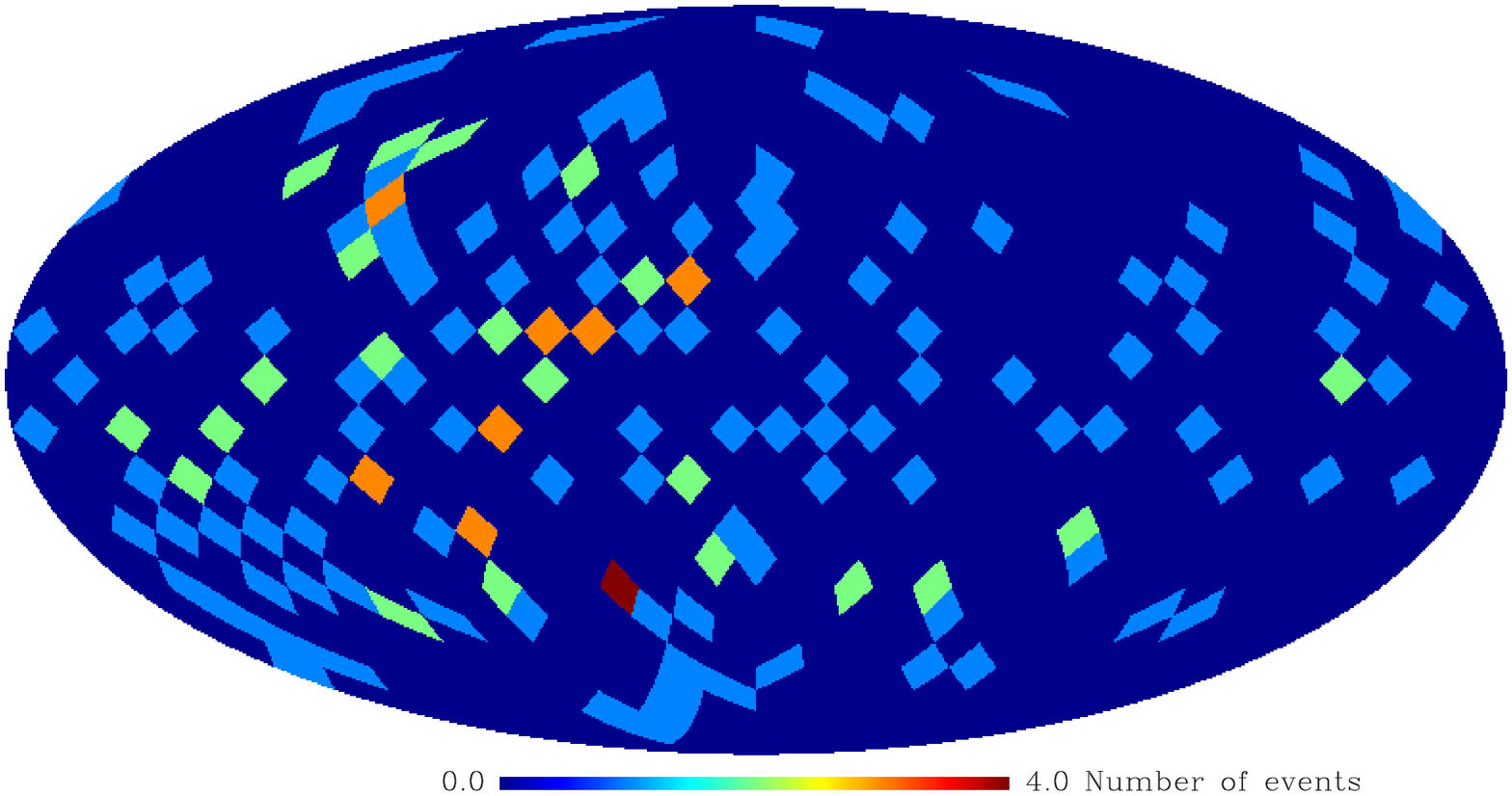}
\caption{From left to right : isotropic background distribution, WIMP-induced recoil distribution  in the case of an isothermal spherical halo  and a typical simulated measurement :  100 WIMP-induced recoils and 100 background events with a low angular resolution. Recoils maps are produced for a $^{19}$F target, a  100 GeV.c$^{-2}$ WIMP  and considering recoil energies in the range  5 keV $\leq E_R \leq$ 50 keV. Figures from \protect\cite{billarddisco}.}  
\label{fig:DistribRecul}
\end{center}
\end{figure}

The main asset of directional detection is the fact that the WIMP angular distribution is pointing towards the Cygnus constellation while  
  the background one is isotropic (fig \ref{fig:DistribRecul}).
The right panel of figure \ref{fig:DistribRecul}  presents a typical recoil distribution observed by a directional detector : $100$ WIMP-induced events and $100$ background events generated isotropically.  
For an elastic axial cross-section on nucleon $\rm \sigma_{n} = 1.5 \times 10^{-3} \ pb$ and a $\rm 100 \ GeV.c^{-2}$ WIMP mass, this corresponds to an exposure of $\rm \sim 7\times 10^3  \ kg.day$ in  $\rm ^{3}He$ and $\rm \sim 1.6 \times 10^3 \ kg.day$  in CF$_4$, on their equivalent energy ranges as discussed in ref. \cite{billarddisco}.
  Low resolution maps are used in this case ($N_{\rm pixels} = 768$) which is sufficient  for the low  angular resolution, $\sim 15^\circ$ (FWHM), expected for this type of detector. In this case, 3D read-out and sense recognition are considered, while background rejection  is based on electron/recoil discrimination by track length and energy  selection \cite{grignonMPGD}.
It is not straightforward to conclude from the recoil map of figure \ref{fig:DistribRecul} (right) that it does contain a fraction of WIMP events pointing towards the direction of the solar motion. 
To extract information from this measured map, 
a likelihood analysis has been developed.
The likelihood value is estimated using a binned map of the overall sky with  Poissonian statistics,  as shown in Billard {\it et al.} \cite{billarddisco}.
This is a four parameter likelihood analysis with $m_\chi$,  $\lambda = S/(B+S)$ the  WIMP fraction ($B$ is the  background spatial distribution taken as isotropic and $S$ is the WIMP-induced recoil distribution) and the coordinates ($\ell$, $b$) referring to the maximum of the WIMP event angular distribution.

\begin{figure}[h!]
\begin{center}
\includegraphics[scale=0.45]{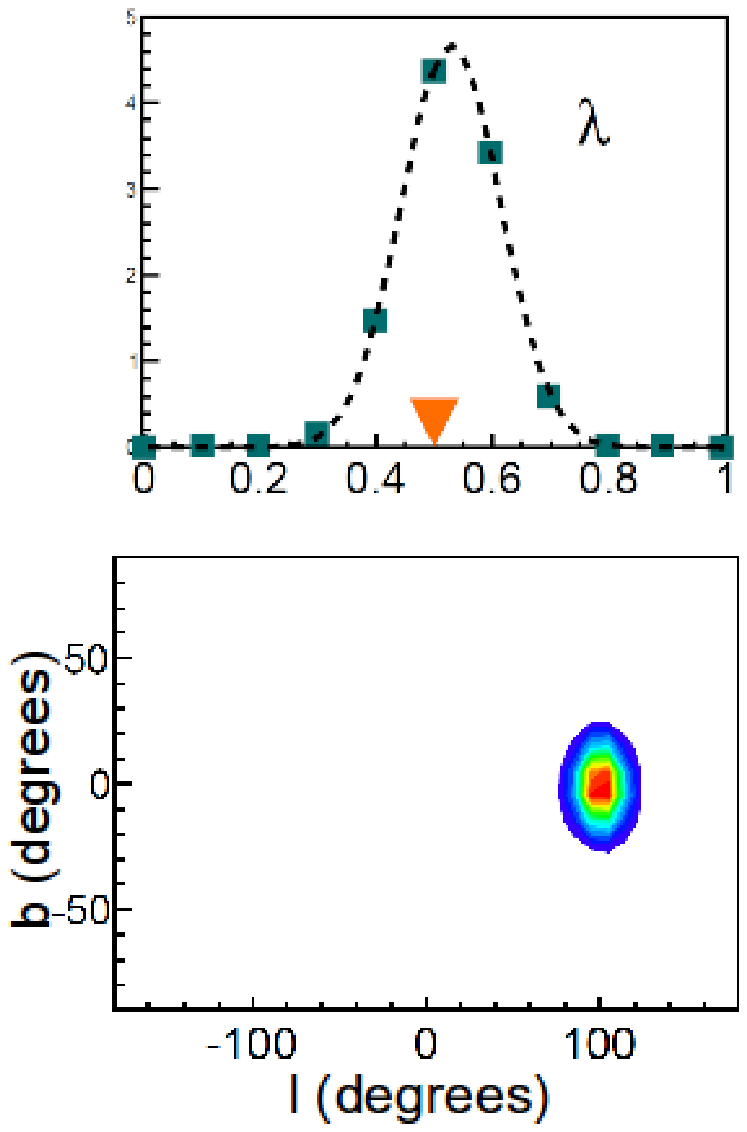}
\includegraphics[scale=0.30]{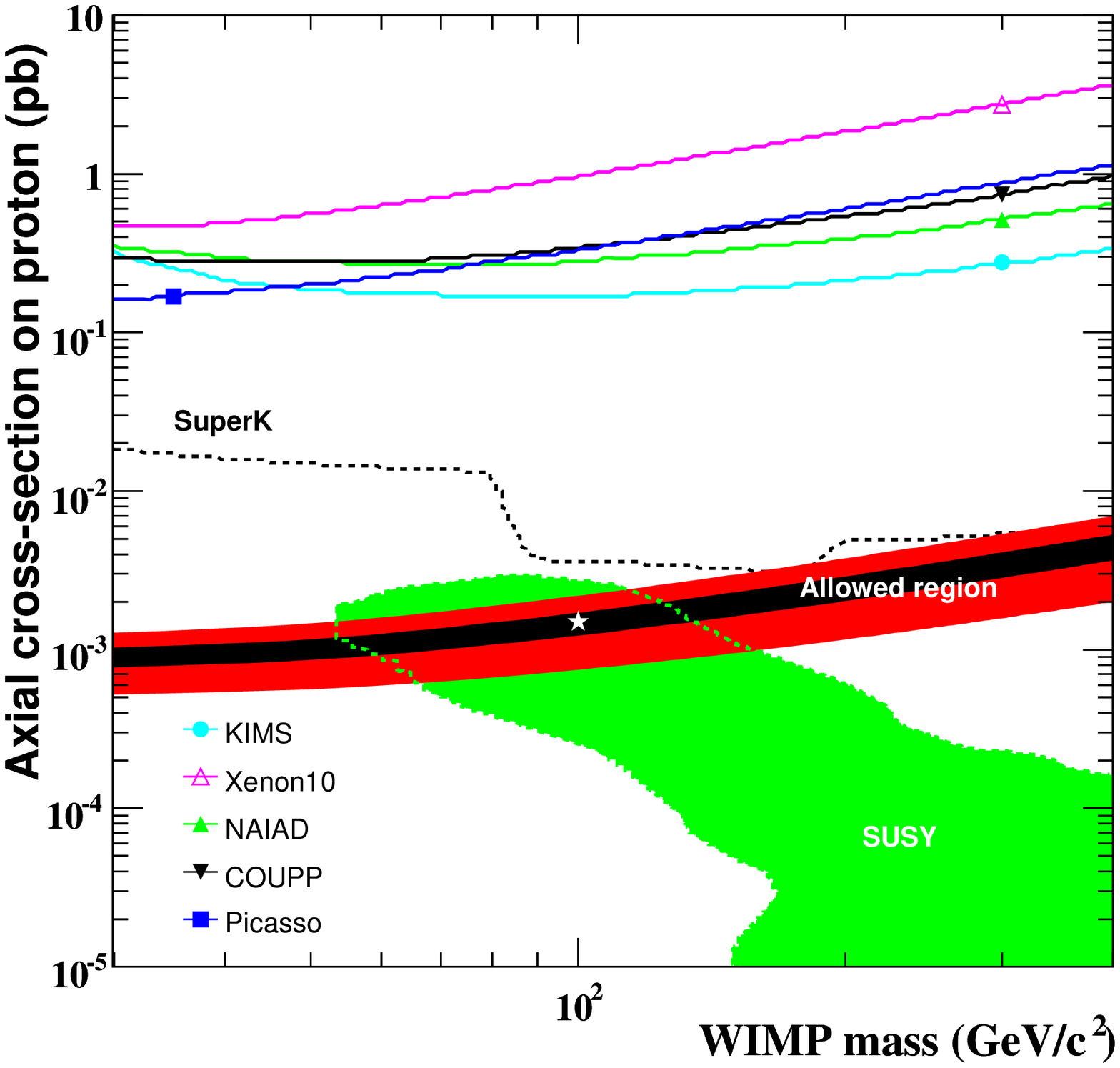}
\caption{On the left, marginalized probability density functions of $\lambda$ (top), $\ell, b$ (bottom) after the likelihood analysis of the simulated recoil map of fig \ref{fig:DistribRecul} right. On the right, allowed regions obtained with the example map shown on figure \ref{fig:DistribRecul} presented in the plane of the spin dependent cross-section on proton (pb) as a function of the WIMP mass (GeV/c$^2$). Input value for the simulation is shown with a star. Figures from \protect\cite{billarddisco}.}
 \label{fig:DiscAndExcl}
\end{center}
\end{figure}

The result of this  map-based likelihood method is that the main recoil direction is recovered and it is  pointing towards ($ \ell = 95^{\circ} \pm 10^{\circ}, b = -6^{\circ} \pm 10^{\circ}$) at $68 \  \%$ CL, corresponding to a non-ambiguous detection of particles from the galactic halo. This is indeed the discovery  proof of this detection strategy (left panel of fig. \ref{fig:DiscAndExcl}) \cite{billarddisco}.
Furthermore, the method allows to constrain the WIMP fraction in the observed recoil map leading to a constraint in the $(\sigma_n, m_\chi)$ plane (right panel of fig. \ref{fig:DiscAndExcl}).  
As emphasized in ref. \cite{billarddisco}, a directional detector could allow for a high 
significance discovery of galactic Dark Matter even with a sizeable background contamination.
For very low exposures, competitive exclusion limits may also be imposed \cite{billardexclu}. However, to use this  very promising strategy with directional detection, access to the recoiling angle of the nuclei is mandatory.

\section{The MIMAC prototype}

In order to get the angle, a 3D reconstruction of a few keV recoil track is needed.
Low-pressure gaseous $\mu$TPC detectors present the privileged feature of being able to reconstruct the track of the recoil following the interaction, thus allowing 
to access both the energy and the track properties (length and direction). 
The MIMAC detector, described in ref. \cite{grignonMPGD}, is based on a matrix of $\mu$TPC containing low mass targets ($\rm ^3He$, $\rm CF_4$, $\rm CH_4$ or $\rm C_4H_{10}$) sensitive to axial interaction.
Each $\mu$TPC uses a micro-pattern gaseous detector (a pixelized bulk micromegas \cite{bulk}) to precisely measure the recoil energy and the track by collecting the ionisation signal.
For a given energy, an electron track in a low pressure micro-TPC is an order of magnitude longer than recoil one. It opens the possibility to discriminate electrons from nuclei recoils by using both energy and track length informations, thus cleaning the recoil dataset before the map analysis presented in section \ref{sec:principle}.
A precise assessment of the energy resolution has been performed by Santos {\it et al.} \cite{santosQuenching} in He + 5\% $\rm C_4H_{10}$ mixture within the dark matter energy range (between 1 and 50 keV) by a precise measurement of the ionization quenching factor.

\begin{figure}[h!]
\begin{center}
\includegraphics[scale=0.65]{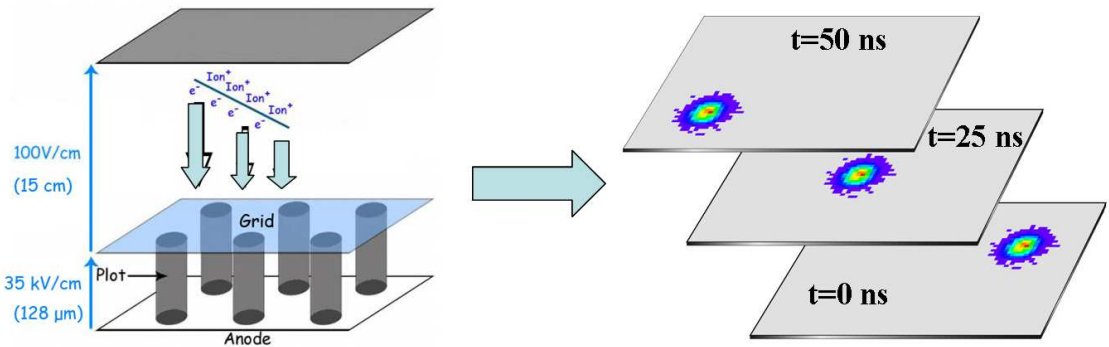}
\caption{Track reconstruction in MIMAC. The anode is scanned every 25 ns and the 3D track is recontructed, knowing the drift velocity,  from the serie of images of the anode.}
\label{recon}
\end{center}
\end{figure}

As pictured on the figure  \ref{recon}, the electrons move towards the grid in the drift space and are projected on the anode thus allowing to get 
information on x and y coordinates. The third coordinate is obtained by sampling the anode every 25 ns and by using the knowledge of the drift velocity of the electrons.
To access the X and Y dimensions with a 100 $\mu$m spatial resolution, a bulk micromegas \cite{bulk}  with a 4 by 4 cm$^²$ active area, segmented in 350 $\mu$m pixels is used with a 2D readout.
 In order to reconstruct the third dimension Z of the recoil, the LPSC developed a self-triggered electronics able to perform the anode sampling at a frequency of 40 MHz.
This includes a dedicated 16 channels ASIC \cite{richer} associated to a DAQ \cite{bourrion}.

\section{First experimental results}

\begin{figure}[h!]
\begin{center}
\includegraphics[scale=0.62]{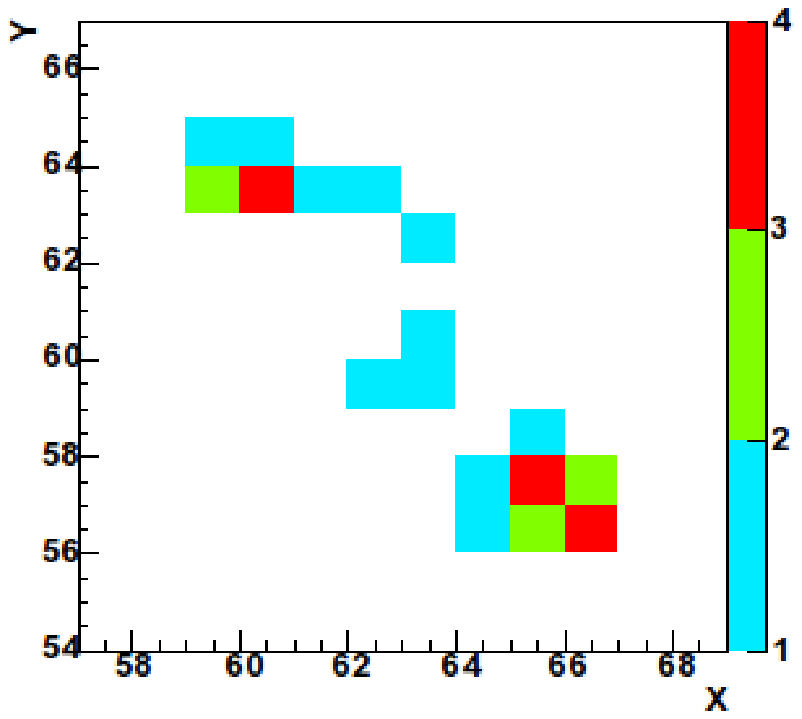}
\includegraphics[scale=0.31]{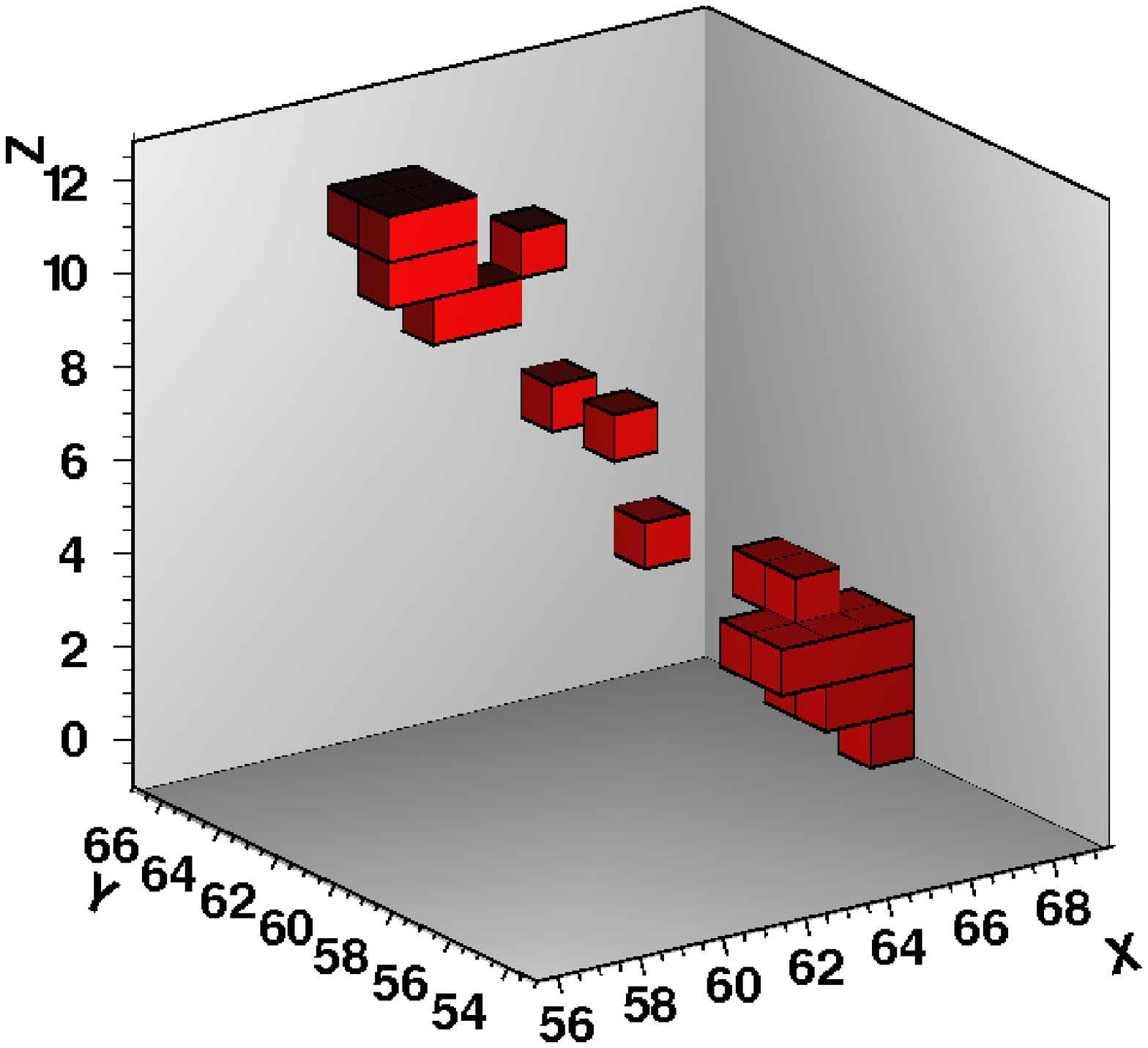}
\includegraphics[scale=0.62]{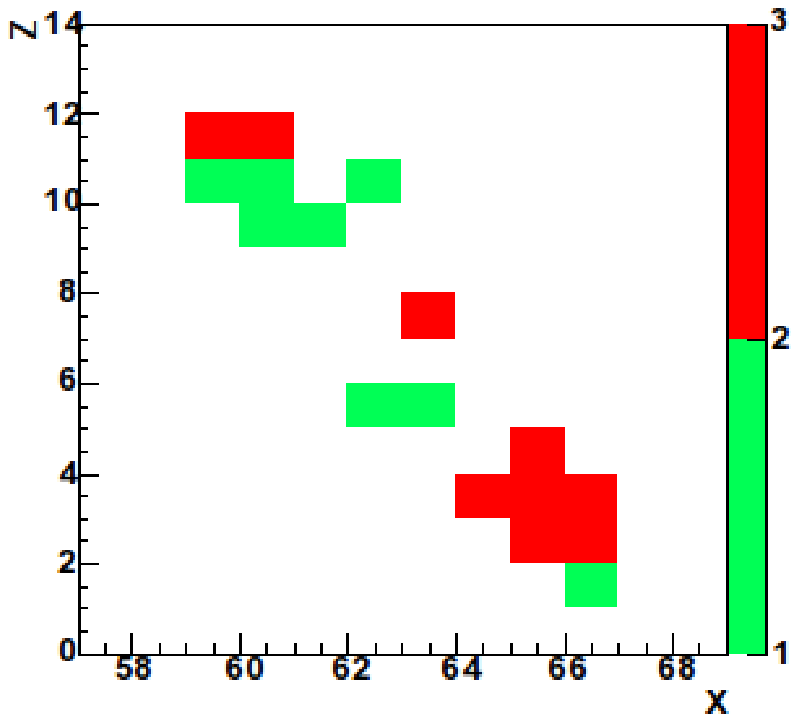}
 \caption{5.9 keV electron track in 350 mbar 95\%~$\rm ^4He+C_4H_{10}$ the left panel represents the 2D projection of the recoil seen by the anode, the center panel represent a 3D view of the track after using the algorithm and the right panel represent a projection of the 3D track on the XZ plane}
 \label{electronTrack}
\end{center}
\end{figure}

First test of the prototype was performed with a $\rm ^{55}Fe$ X-ray source in order to reconstruct 5.9 keV electrons.
Figure \ref{electronTrack} presents a typical electron track seen by the anode (left panel), reconstructed in 3D (center panel) and projected on the XZ plane (right panel).
This result shows the MIMAC capability to reconstruct and identify low energy electrons which are the  typical background in dark matter experiments.

\begin{figure}[h!]
\begin{center}
\includegraphics[scale=0.32]{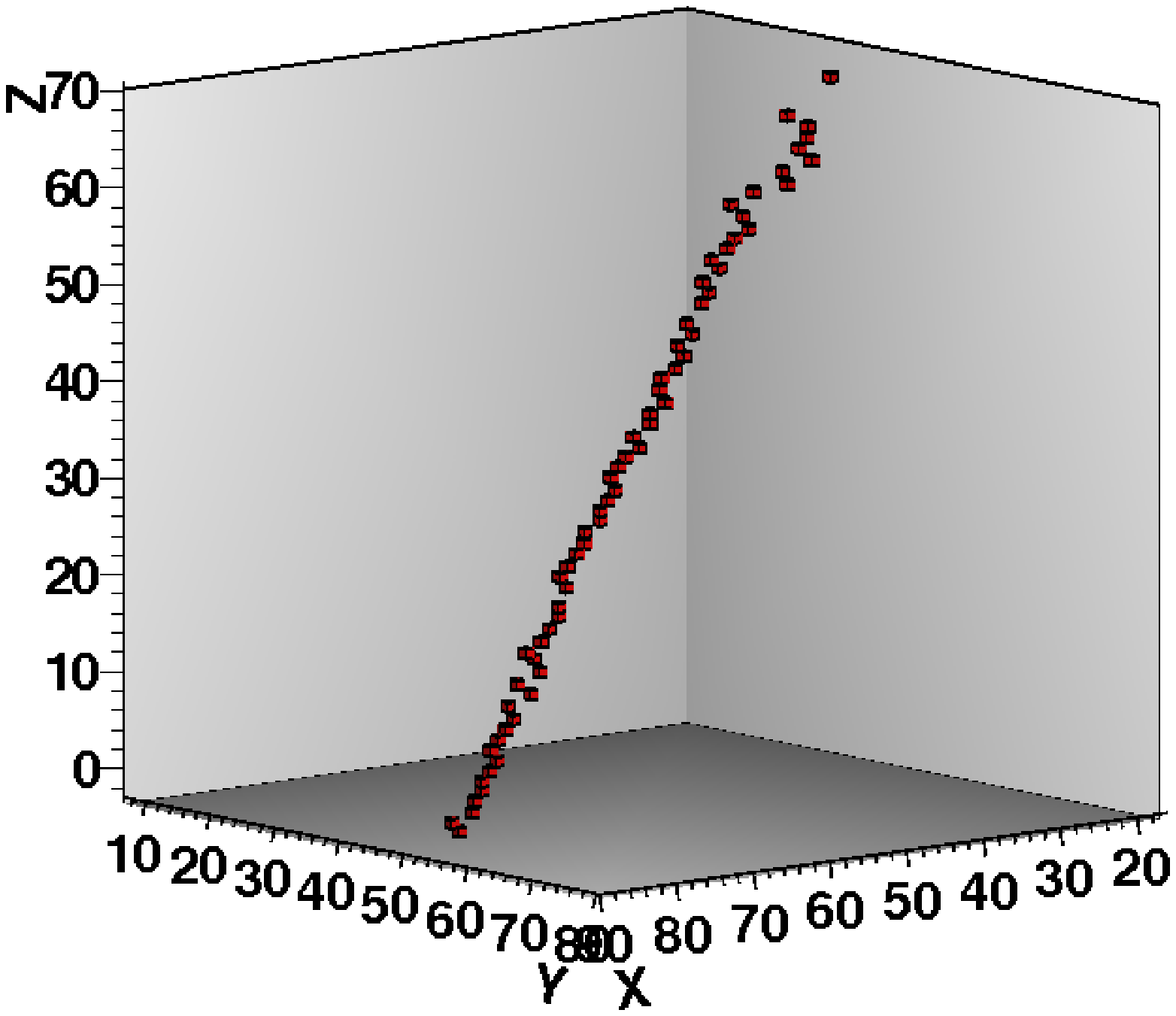}
\includegraphics[scale=0.32]{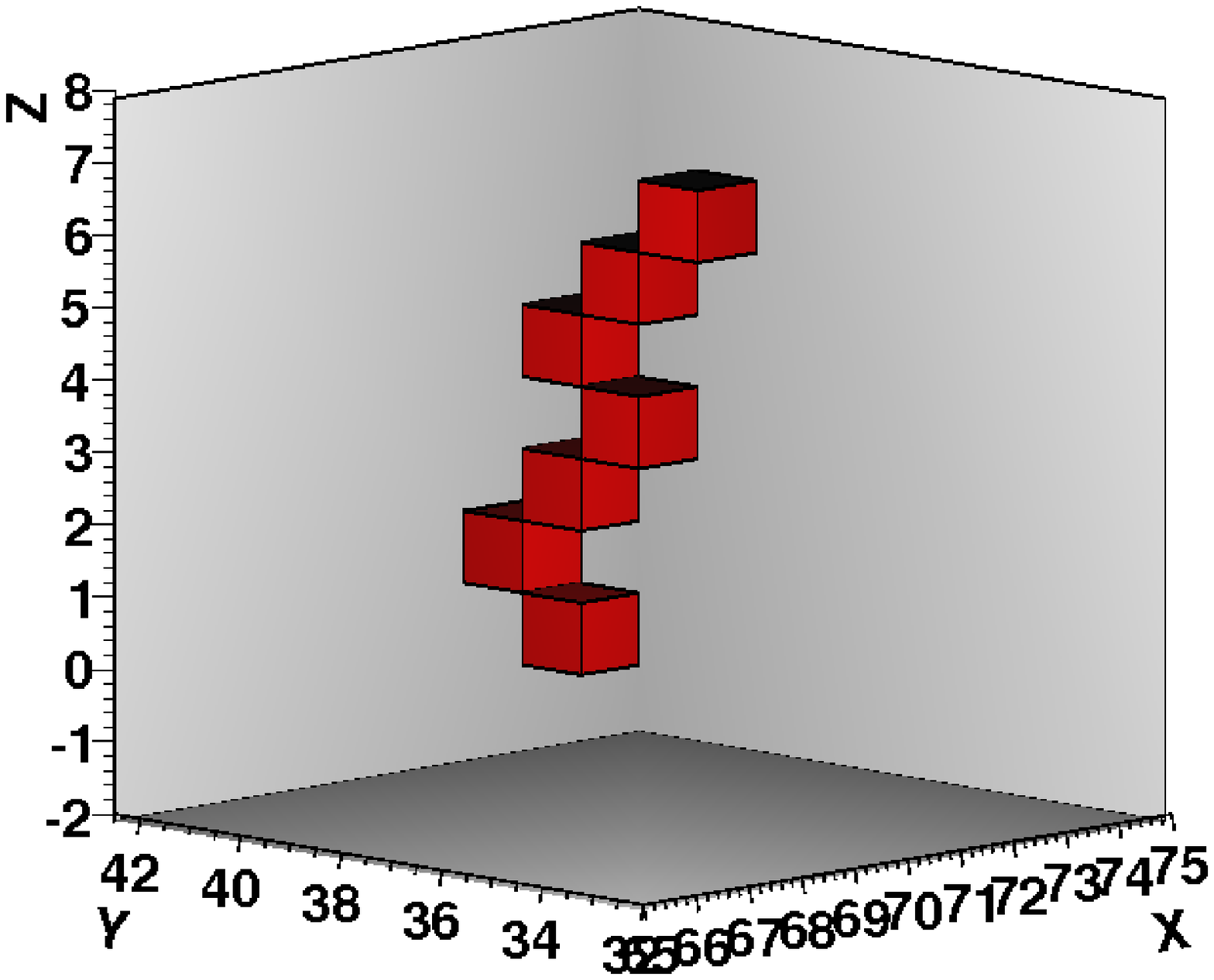}
\includegraphics[scale=0.32]{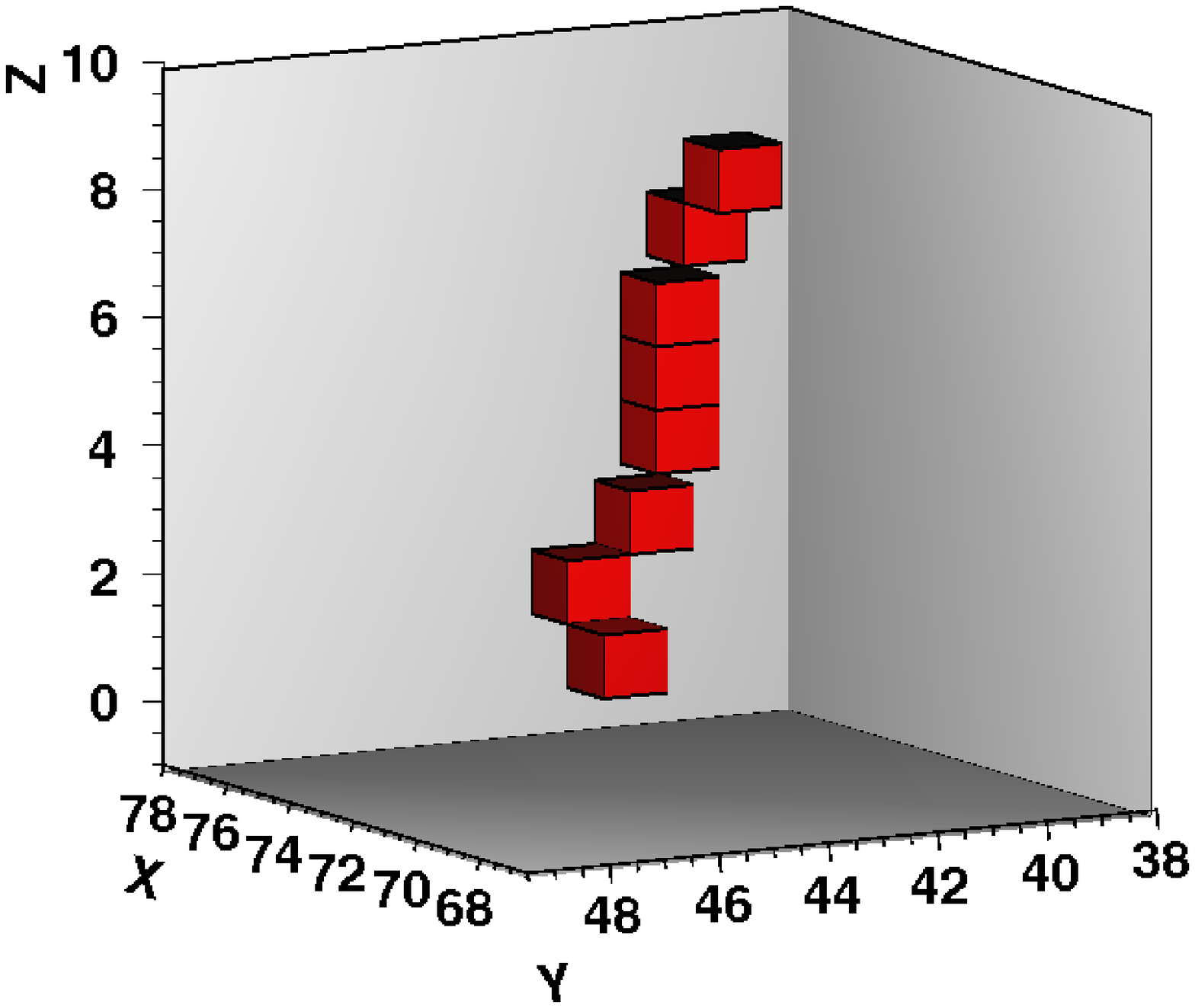}
\caption{From left to right: recoil of 5.5 MeV He nuclei  in 350 mbar $\rm ^4He+ 5\% C_4H_{10}$,  8 keV hydrogen nuclei in 350 mbar $\rm ^4He+ 5\% C_4H_{10}$ and 50 keV fluorine nuclei in 55 mbar 70\% $\rm CF_4$ + 30\% $\rm CHF_3$}
\label{nucleiTracks}
\end{center}
\end{figure}

As shown on fig \ref{nucleiTracks}, 3D reconstruction is achieved for high energy (few MeV) alpha particles issued from the natural radioactivity ($\rm ^{222}Rn$) present in the gas.
On the same figure, 3D tracks on nuclei recoils following elastic scattering of mono-energetic neutrons are represented.
On the center panel, a 8 keV proton recoil of 2.4 mm long in 350 mbar $\rm ^4He+ 5\% C_4H_{10}$ is represented: it is the typical kind of signal that MIMAC will expect.
The right panel presents a 50 keV (with quenching) fluorine recoil of 3 mm long obtained in a 55 mbar mixture of 70\% $\rm CF_4$ + 30\% $\rm CHF_3$: the preliminary result is very promising for future directional detectors as the one simulated to obtain the exclusion plots presented in part \ref{sec:principle}.
Furthermore, another important point for dark matter detection, the ability to separate of gamma background from WIMPs, has been shown in pure Isobutane or $\rm ^4He+ 5\% C_4H_{10}$ mixture in ref. \cite{grignonMPGD}.

\section{Conclusions}

Directional detection is a promising search strategy to discover galactic dark matter.
The MIMAC detector provides the energy of the recoiling nuclei and the reconstruction of the 3D track. 
Firsts 3D tracks were reconstructed with the MIMAC prototype: 5.9 keV electrons (typical background) and low energy proton and fluorine recoils (typical signal).

\end{document}